# Intelligent machines work in unstructured environments by differential neuromorphic computing

**Authors**

Shengbo Wang[†1], Shuo Gao[†*1], Chenyu Tang[2], Edoardo Occhipinti[3], Cong Li[1], Shurui Wang[1], Jiaqi Wang[1], Hubin Zhao[4], Guohua Hu[5], Arokia Nathan[6], Ravinder Dahiya[7], Luigi Occhipinti[*2]

**Affiliations**
[1]School of Instrumentation and Optoelectronic Engineering, Beihang University, Beijing, China
[2]Department of Engineering, University of Cambridge, Cambridge, UK
[3]Department of Computing, Imperial College London, UK
[4]HUB of Intelligent Neuro-engineering (HUBIN), CREATe, Division of Surgery and Interventional Science, UCL, HA7 4LP, Stanmore, UK
[5]Department of Electronic Engineering, The Chinese University of Hong Kong, Shatin, N. T., Hong Kong S. A. R., China
[6]Darwin College, University of Cambridge, Cambridge, UK
[7]Bendable Electronics and Sustainable Technologies (BEST) Group, Department of Electrical and Computer Engineering, Northeastern University, Boston, MA 02115, USA
[†]These authors contributed equally to this work
[*]Correspondence to: shuo_gao@buaa.edu.cn, lgo23@cam.ac.uk

**Abstract**

Efficient operation of intelligent machines in the real world requires methods that allow them to understand and predict the uncertainties presented by the unstructured environments with good accuracy, scalability and generalization, similar to humans. Current methods rely on pretrained networks instead of continuously learning from the dynamic signal properties of working environments and suffer inherent limitations, such as data-hungry procedures, and limited generalization capabilities. Herein, we present a memristor-based differential neuromorphic computing, perceptual signal processing and learning method for intelligent machines. The main features of environmental information such as amplification (>720%) and adaptation (<50%) of mechanical stimuli encoded in memristors, are extracted to obtain human-like processing in unstructured environments. The developed method takes advantage of the intrinsic multi-state property of memristors and exhibits good scalability and generalization, as confirmed by validation in two different application scenarios: object grasping and autonomous driving. In the former, a robot hand experimentally realizes safe and stable grasping through fast learning (in ~1 ms) the unknown object features (e.g., sharp corner and smooth surface) with a single memristor. In the latter, the decision-making information of 10 unstructured environments in autonomous driving (e.g., overtaking cars, pedestrians) is accurately (94%) extracted with a 40×25 memristor array. By mimicking the intrinsic nature of human low-level perception mechanisms, the electronic memristive neuromorphic circuit-based method, presented here shows the potential for adapting to diverse sensing technologies and helping intelligent machines generate smart high-level decisions in the real world.

**Introduction**

Understanding and learning sensory data efficiently to achieve human-like perception of the real world is pivotal for next-generation intelligent machines (*1–3*). With such capabilities, intelligent machines could truly transit from controlled environments such as factories and laboratories into unstructured environments of home and businesses that offer considerable 'variations'. Traditionally, the adeptness of organisms within unstructured environments has been attributed to the acquisition of diverse types of physical information. However, recent advancements in life sciences have revealed that the most crucial mechanism employed by humans for understanding unstructured environments is sensory information differentiation and learning mechanisms (*4–8*). For a given stimulus, multiple types of receptors and corresponding neurons participate in differential processing, and they adjust their structure and synaptic weight on the features of external stimuli. Specifically, by extracting the main stimulus features embedded in the signal properties, these receptors and neurons can rapidly form a complex set of intricate network-based perception functions, such as integration and firing, potentiation and depotentiation, associative memory, environmental mapping, and motion control tasks. In contrast, conventional intelligent machines operate within the von Neumann separated storage and computation architecture, which makes it inefficient to differentially process highly dynamic unstructured information (*9–13*). This can be resolved with memristors, as they integrate storage and computation capabilities, coupled with synapse-like characteristics, and thus bear a striking resemblance to the biological synapse at a foundational level (*14–19*). This similarity opens the possibility for the realization of human-like perception functionalities in intelligent machines. To this end, herein we present a memristor-based differential neuromorphic computing, perceptual signal processing and learning method for intelligent machines.

Unstructured information encompasses multidimensional features, which undergo the perceptual processing of diverse receptors in biology (*2, 20, 21*). This type of processing indicates the need for designing different modulation methods for neuromorphic computing. However, existing methods focus on keeping a memristor to a fixed receptor (Figure 1a), e.g. a nociceptor (*22–25*), thus restricting their full potential. A potential solution is to operate multiple memristors at different desired states. In fact, the adjustable characteristics of memristors allow them to switch their states, which could be used to focus on multiple receptors. Furthermore, distinct features of sensory data exhibit time domain independence, implying that a memristor can adaptively switch to an appropriate state to process each feature at a time slot. Therefore, drawing inspiration from the sensory information processing model and taking advantage of memristors' intrinsic multistate property, the proposed differential information processing and learning method involves extracting features from unstructured data and applying memristive modulation schemes. This approach experimentally facilitates the processing and learning of unstructured data within two complex environments i.e., grasping of objects and autonomous driving. In the former, we emulate nociceptors to achieve amplification (>720%) for hazardous stimuli and adapting receptors to achieve adaptation (<50%) for mild stimuli, which play a key role in grasping unknown objects. In the latter, we differentiate visual information by a 40×25 memristor matrix and reach a commendable accuracy of 94% in extracting critical information within 10 autonomous driving scenes, after comparison with human labeled results.

## Results

### Specific Components of Differentiation Processing and Environmental Learning

Sensory, signal encoding and neuromorphic operation modules construct the differential neuromorphic computing method (Figure 1b). The first can be of the desired sensor type according to mimicked biological sensory. It operates by converting a physical stimulus into electrical information and can be described as:

$$p_i(t) = R_i(s_i(t)) \tag{1}$$

where $s(t)$ denotes the physical stimulus, $R$ is the response function of the sensor, and $p(t)$ is the outputted electrical signal for each $i$ th channel/memristor.

The signal encoding module extracts features from $p(t)$ and then creates the associated memristive encoding schemes to process sensor information in different manners. The created encoding scheme is applied to the memristor in the neuromorphic operation module, and the changed resistive values indicate the properties of the suffered stimuli. Through the three steps above, different features are properly processed by the predesigned differential modulation methods for memristors, yielding a multifeature differentiation-based comprehensive understanding of environmental knowledge. The entire process corresponds to organisms' differential information perception capability in stimuli reception, transduction and processing (26), as expressed below:

$$f_i(t) = F_i(p_i(t)) \tag{2}$$

$$v_i(t) = V_i(f_i(t), M_i(x_i(t-1))) \tag{3}$$

$$\frac{dx_i}{dt} = C_i(x_i, v_i) \tag{4}$$

where $F$ is the extracting function, $f(t)$ is the current extracted features, $x(t-1)$ is the memristor state after the previous modulation, $M$ is the eigenvalue used for the memristor weight update and associated with its state, $V$ is the piecewise memristive encoding scheme responsible for generating current modulation signals $v(t)$, $\frac{dx}{dt}$ is the derivative of the memristor state variable, and $C$ is the state derivative function related to the memristor mechanism, current state and external modulation voltage for each $i$ th channel/memristor.

As explained, the proposed method provides a human-like information processing pipeline, which extracts key features of undergoing stimuli in real-time, opening the possibility for intelligent machines to learn the main attributes of unstructured tasks automatically, as demonstrated in the following tactile and visual information processing experiments.

### Tactile Perception

Realizing safe and stable manipulation of unknown objects by robotic hands is tricky but highly needed (27–31). Unlike lab and factory settings, an unknown object may exhibit sharp or slippery attributes. The former potentially damages the contacting end, and the latter places a heavy burden on the sensing and computing modules when keeping the object balanced between stabilization and deformation. To address this issue, the proposed memristor-based differential neuromorphic computing method is embedded into the sensing and controlling strategy of a robotic hand, as depicted in Figure 2a.

Here, a piezoresistive force sensing architecture (more details in Figure S1) is assembled for receiving pressure amplitudes, and a self-directed channel effect-based multilayered nonvolatile memristor (KNOWM Inc.) is selected for differential neuromorphic computing. The structure, hysteresis curves, and electrical characteristics under pulse testing of the memristor are given in Figures 2b to d, respectively. The piezoresistive layers and the memristor offer short-term and long-term force information, respectively, based on which the attributes of an object are first extracted by a status acquisition block and then utilized to generate differential modulation schemes for the memristor. The extraction and modulation scheme generation are both conducted in a field programmable gate array (FPGA) platform, as shown in Figure 2a. Based on the aims of modulation methods, the schemes can be further classified into 3 groups, allowing the memristor to stay at high (>250k), middle (~170k) and low (<100k) resistance levels. This arrangement facilitates the adaptive, normalization, and nociception perception of external stimuli, aligning with 3 types of biological perception behaviors, i.e., nociception, adaptation, and recovery. The detailed mapping relationship between attributes and modulation methods is given in the Supplementary Table 1.

The three mimicked biological functionalities are experimentally validated, by studying the change in force induced electrical signals with and without neuromorphic computing. Here, the neuromorphic computing module tunes the adjustment factor of a voltage amplifier block, whose input is the status of the piezoresistive film. After the attribute of a force is extracted, an associated modulation scheme is selected and then applied to the memristor. The resistance of the memristor is then changed and used as an indicator of the amplification factor. The final yielded voltage outputs illustrate the biological perception behaviors. The experimental results in Figure 2f and h demonstrate the responses of neuromorphic computing to hazardous and mild stimuli. More than 170% amplification of the hazard signal and more than 50% attenuation of the mild signal are obtained. The curve trends are similar to the biological response strengths in Figure 2g.

When considering the unstructured information processing in the real world, it often becomes necessary to perform further differentiation on the aforementioned functions. For better adaptation to dynamic changes in hazardous scenarios, we mimic the time window processing mechanism in organisms, thereby distinguishing between sudden threats and persistent threats (*24*, *32*). Specifically, during the encoding process, we additionally consider the state of the memristor. When its state falls below a certain threshold (100 kΩ in this case), we increase the amplitude and pulse duty cycle of the positive pulses (as shown in Figure 2i), achieving the functionality of sensitized processing for external stimulus information. In our work, a >720% amplification of tactile stimuli is offered (Figure 2g), exceeding the state-of-the-art works (*22*, *25*, *33*). Regarding the adaptation function, the adaptive speed (the magnitude of the attenuation of response strength over time) is of significance (*34–36*). Here, we implement two adaptation speeds by adjusting the pulse duty cycle and amplitude of encoding pulses as shown in Figures 2k and 2l. The developed method is then used in two practical unknown object grasping tasks, to extract and learn the main characteristics of objects for safe and stable manipulation (conceptually depicted in Figure 3a).

In task 1, an irregular object consisting of a cube and a cone is 3D printed. During the grasping process, the robot might encounter the sharp point that leads to painful pressures and poses a potential danger. It needs the help of a nociceptor to quickly learn the hazard stimulus caused by the sharp point and then changes the grasping strategy, as conceptually depicted in Figure 3b, in which the resistance information of the piezoresistive film and the memristor during the entire grasping task is shown. At 2.2 s, the piezoresistive sensor attached to the robotic hand experiences a significant force due to contact with the sharp point, resulting in decreased resistance. In this case, the pressure conforms to hazardous characteristics, and the memristor is modulated into the amplification state (low resistance). At 8.2 s, the memristor resistance falls below the threshold of 35 k$\Omega$, and the pressure stimulus is amplified by 500%. This indicates that the robot has been in contact with the sharp point, triggering the pain reflex. Subsequently, the robot learns to change the grasping posture at 14.4 s. By 17.4 s, the pressure perceived by the contact point stabilizes, and the memristor is modulated into an adaptation state (high-resistance state), achieving safe and stable grasping.

Similarly, in task 2, a soap is used to represent slippery objects. The instability of tactile information during the slipping process leads to spikes in reading the resistance of the memristor. The spike signals due to slip events prompt the robot to increase the gripping force to ensure stability. As illustrated in Figure 3c, during the grasp attempt of the soap, the gripping force stabilizes at 3.8 s, with the memristor switching to a high-resistance state. By 18 s, the adaptation level to external tactile information reaches 75%. At 18.5 s, an external inference causes the object to slip, leading to changes in gripping force and resulting in a spike in the memristor resistance. Upon detecting a slip event onset, the robot prompts an increase of the gripping force to prevent the object from slipping further. This timely increase in contact force prevents the object from falling, and the memristor is modulated into a normal perception state (stable middle resistance state), successfully achieving stable grasping of a slippery object. To our knowledge, this represents the first time that a memristor-based approach is successfully employed for local slip detection and consequent adjustment of the actuator force when grasping unknown objects.

At present, the processing and execution time of the proposed method is 1 ms, which can strongly support safe and stable operations for robots (*37–40*). The operation time can be further improved when better memristive devices that respond in nanoseconds (ns) or microseconds (*μ*s) are employed (*41–44*).

**Visual Perception**

In previous tactile information processing, attributes based on force strength were used to learn environmental knowledge. In contrast, visual frequency-based attributes are more important for autonomous driving, as they imply the relative position change of surrounding objects to the car (*45–47*) and are vital for real-time decision making (*48–50*). For instance, the sudden appearance of pedestrians or vehicles can lead to life-threatening collisions. Thus, shortening the attribute extraction time of moving objects is important. In addition, the moving direction of the object can further help to make good decisions (*51, 52*). To this end, the proposed differential neuromorphic method is employed to learn visual frequency-based attributes, forming functions mimicking biological cone cells for retrieving fast information, and neural excitation to maintain fast information for a while.

In the implementation, a driving recorder of 1920×900 resolution and a 25×40 memristor array are used. The gray CMOS image from the recorder is compressed, by averaging a matrix of m×n into one pixel, based on image spatial redundancy theory (*53*). Next, filtering circuits are used to extract fast and slow pixel value changes within two adjacent frames, obtaining fast and slow visual information attributes, which are then encoded into electrical pulses to modulate the memristor into low- and high-resistance states. Here, slow visual information is taken to release the memristor from a low-resistance state to a high-resistance state. As the change in resistance is analogue, its value within this period reflects the moving orientation of the object, achieving a neural excitation effect.

In the experiments, the driving operations are divided into designed driving and free driving scenarios. In the former, slow driving takes place on a closed road segment, and a pedestrian runs across the road from different sides (left and right), distances (near, medium and far) and speeds (walking and running), aiming to examine the detection ability of the method in 3 widely occurring danger scenarios for autonomous driving as shown in Figure 4b: 1. Pedestrian running across the road (moment 1); 2. Nearby pedestrian walks across the road (moment 2); 3. The walking pedestrian (moment 3) suddenly runs to another orientation (moment 4). Figure 4c displays 3 representative scenarios during a pedestrian's moving path given in Figure 4b, in which the yellow boxes are the m by n pixel area that is first compressed to a single pixel and then processed by the same memristor. The light intensity changes of the pixel are shown in Figure 4d. Before moment 1, no pedestrian enters, and the slight light intensity change is solely due to the vehicle movements. Upon reaching moment 1, the pedestrian runs into the area from a medium distance, resulting in strong light intensity change, triggering the extraction and encoding module to generate a positive modulation voltage pulse, thus modulating the corresponding memristor into a low-resistance state. After the pedestrian leaves the area, the corresponding memristors enter the neural excitation status and finally return to the high resistance status because the scene information changes slowly. At moment 2, the pedestrian walks into the yellow box area, giving rise to a strong light intensity change again. Note that although the pedestrian is in a slow-motion mode, the distance between the vehicle and pedestrian is short, hence showing the same effect as a pedestrian running from far. At moment 3, the pedestrian enters the detection area from far, and the slow movement characteristics generate a negative voltage pulse that maintains the memristor in a high-resistance state. At moment 4, the sudden movement of the pedestrian strongly changes the light intensity, hence reducing the resistance value of the memristor. However, due to the distance of the pedestrian and the fact that he has already entered the area before the sudden run, the amplitude of the positive pulse generated by the high-frequency feature is smaller compared to the previous more dangerous moments (moments 1 and 2). For this reason, the resistance value of the memristor does not reach to its lowest resistance state. Overall, the 3 danger scenarios are successfully detected.

When the yellow box expands to the whole image, global attributes are gathered. Figures 4e to 4g are 3 examples taken from daytime and evening, together with their corresponding differential neuromorphic computing results. The fast-running pedestrian is accurately captured, and afterimages are generated, implying the orientation.

In free-driving experiments, we collected over 100 hours of video data in various lighting and weather conditions, as conceptually shown in the middle part of Figure 5. Ten representative scenarios containing important decision-making associated information for autonomous driving, such as taillight, nearby cars and lane lines, are given as examples in Figure 5. From the differential neurocomputing processed results, it is clear that the information is successfully retrieved. We then further invite 10 senior drivers to locate dangerous objects they believe are important to safe driving in 10 short videos containing the 10 images given in Figure 5. Their results are offered as the yellow regions in the processed results, and we can learn that human decisions overlap with the proposed method. For all 10 videos, the overlap rate is more than 94%. The difference arises from the process of important information labeling. The judgment criterion of manual labeling, based on senior drivers' driving/riding experience, tends to prioritize targets (such as pedestrians and cars) that are likely to cause traffic accidents. In contrast, the criterion of differential processing is based on objective motion situation, which is of greater significance for timely decision-making, assigning relatively stationary or slow-moving objects low priority on current driving. Therefore, when both empirical and objective moving targets are present, a person tends to label empirical targets. However, at times, such as when driving with a vehicle traveling at a relatively close speed nearby, the road marking line is more important and can reflect the yawing condition of driving (Figure 5b). In most cases the empirical and objective moving targets are the same object, but there are exceptions such as in Figure 5h, leading to a 6% bias in the detection results.

Compared to traditional object detection methods in autonomous driving, the developed differential neurocomputing framework does not rely on prior trained dataset, illustrating its great power in processing unstructured information and hence prompt the implementation of real-time decision-making for autonomous driving.

**Conclusion and outlook**

Memristors, with their ability to efficiently mimic the computation units of biology, have been used to construct human-like environmental perception systems (*54*, *55*). However, current memristive computing methods overlook the characteristic attributes of biological information processing (*20*, *22*, *56*, *57*), which could lead to information loss in unstructured environments. Drawing inspiration from biology, we demonstrated the memristor based processing of environmental information by exploiting their diverse characteristics and using multiple receptors. Subsequently, we demonstrate the effectiveness and versatility of this method by showcasing its application in both tactile and visual perception. The presented results represent a significant advancement toward achieving fully autonomous and intelligent robot systems capable of effectively operating in real-world scenarios. Due to the small size of memristive devices, the possibility of having high-density memristors over large areas and flexible substrates, and their similarity with the fundamental biological sensory processing mechanisms, the presented method could enable intelligent machines to possess sensory capabilities on a scale comparable to humans when combined with diverse sensors, allowing them to genuinely learn and efficiently comprehend the environment (as depicted in Figure S22 and S23).

# Figures

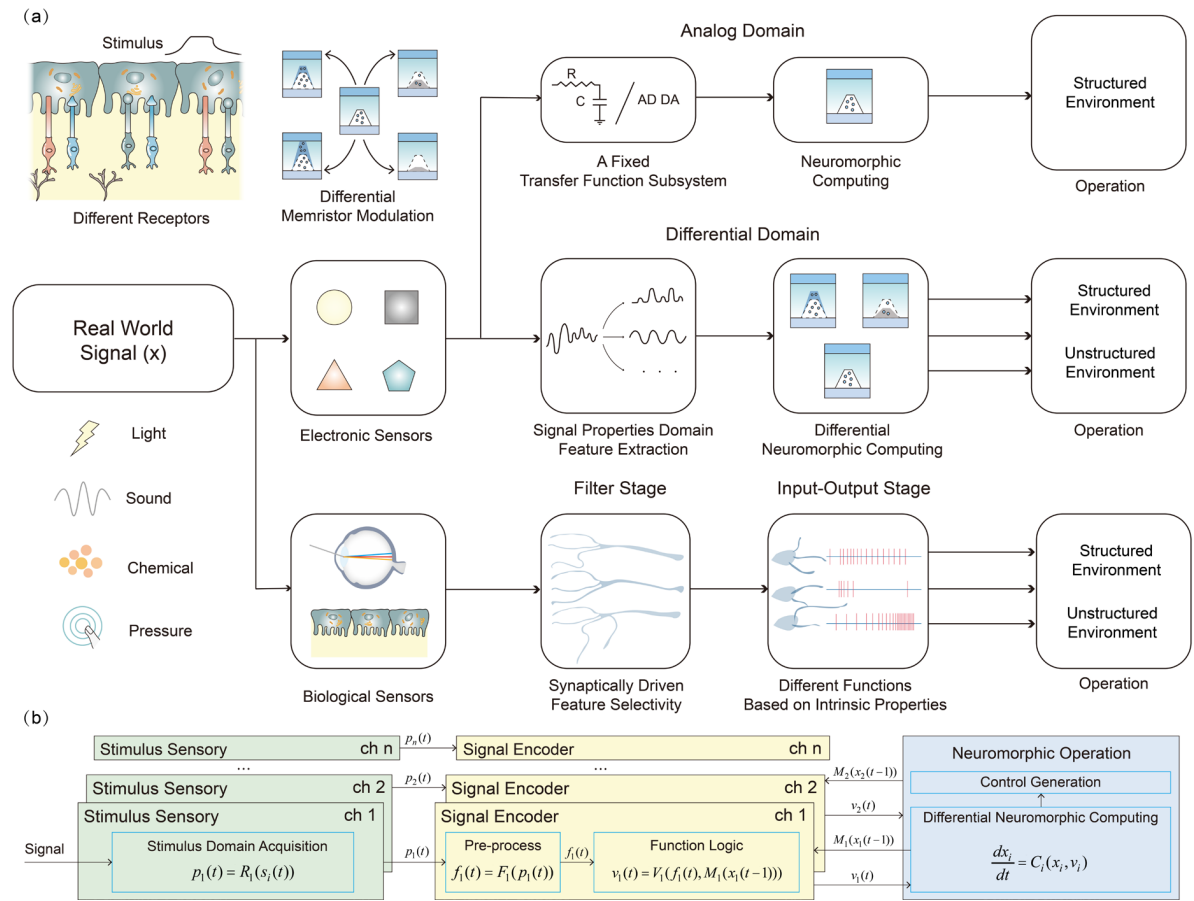

**Figure 1. The proposed differential neuromorphic computing method.** (a) Comparison of our proposed differentiated processing method with biological sensory processing methods and current neuromorphic processing methods. (b) The specific implementation of the memristor-based differentiation processing method.

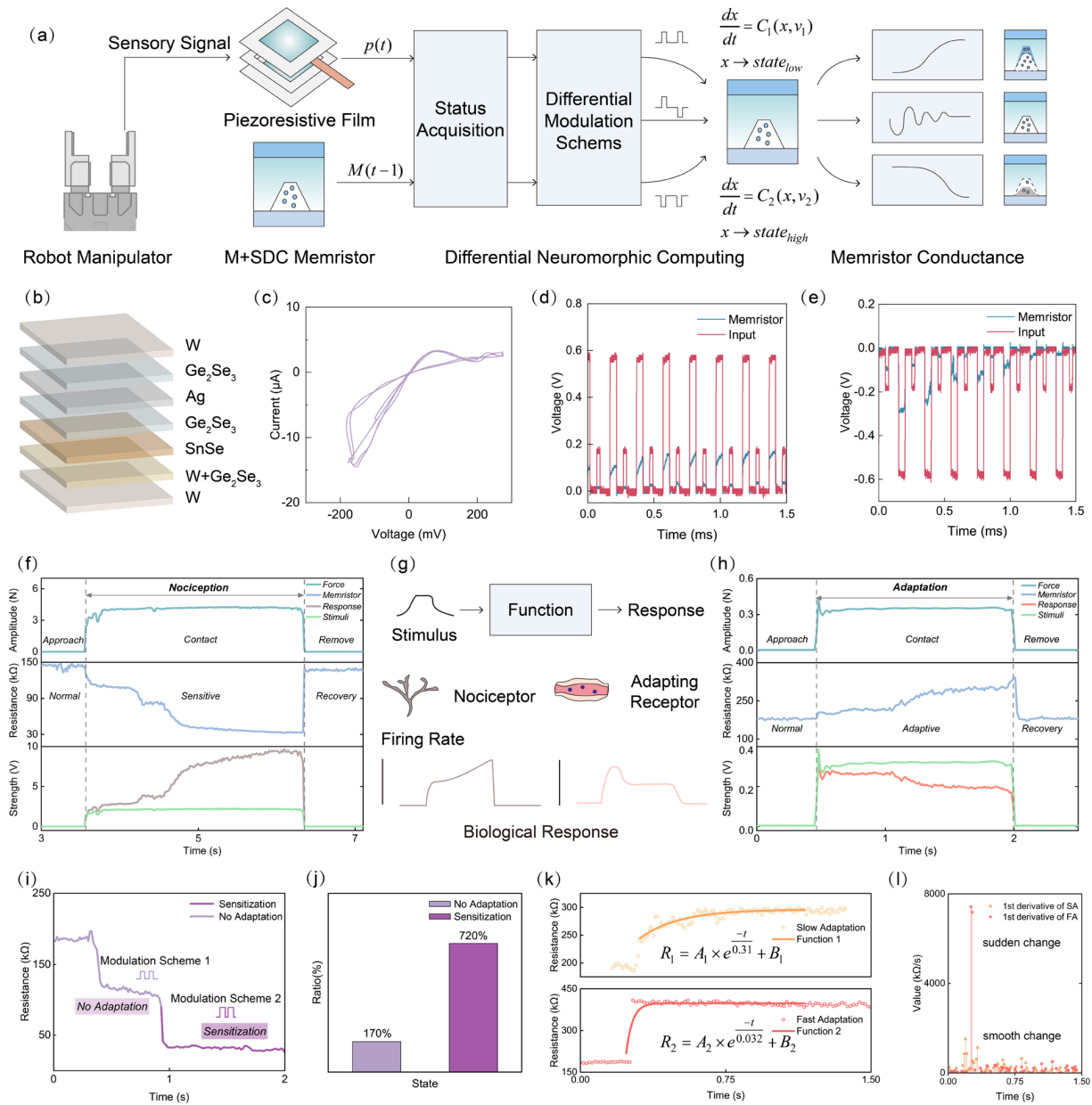

**Figure 2. Utilizing differentiation processing method to process tactile information and mimic multiple receptors**. (a) Schematic diagram of tactile information processing. (b) The device structure of the self-directed memristor. (c) The U-I test of memristors. In this test, a 500mV peak-to-peak sine wave is applied to the memristor and the 10kΩ fixed-value resistor. (d) Electrical characterization of memristors using positive pulses. (e) Electrical characterization of memristors using negative pulses. (f) Amplification processing of hazardous stimuli. (g) Corresponding biological processing functions. (h) Adaptation processing of mild stimuli. （i） Sensitization processing achieved through further differentiation. (j) Quantitative evaluation of the sensitization function. (k) Realization of adaptation at different speeds. (l) The adaptive rate.

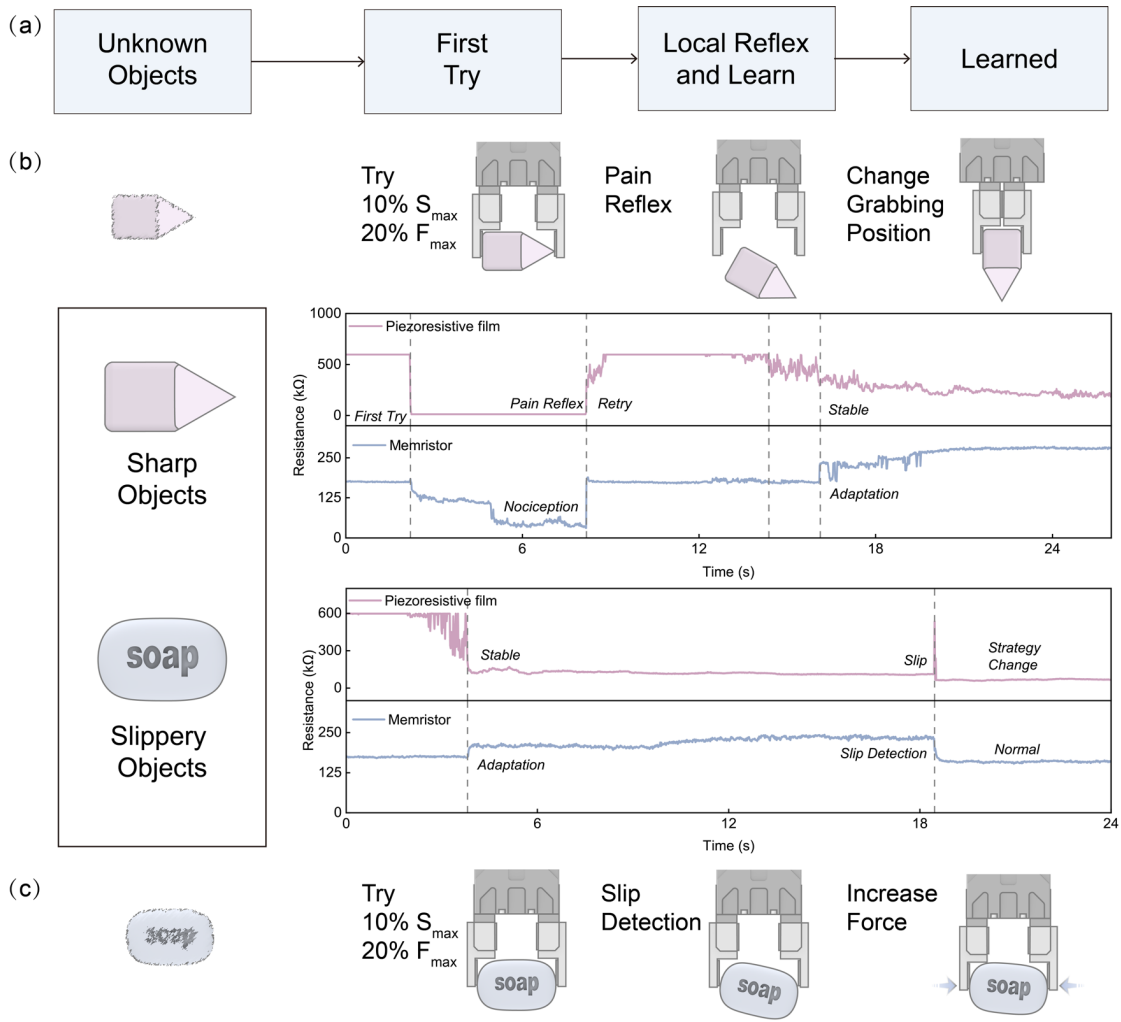

**Figure 3. Grasping diverse objects in unstructured environments.** (a) Overall logical flowchart of learning environmental information. (b) Learning the properties of sharp objects and ultimately achieving safe grasping. (c) Learning the properties of slippery objects and ultimately achieving stable grasping.

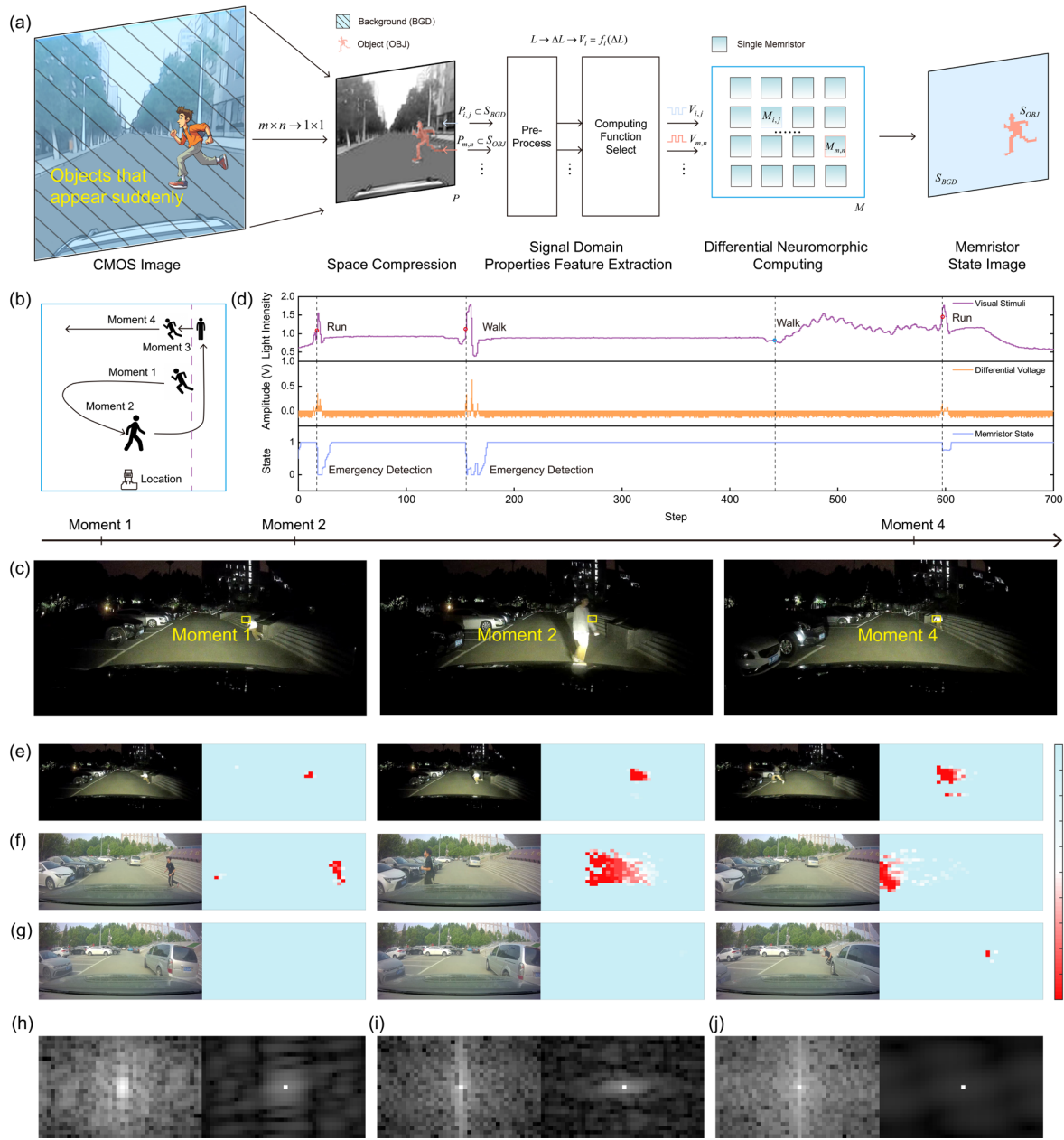

**Figure 4. Utilizing differentiation processing method to process and learn visual information.** (a) The diagram of visual information differentiation processing. (b) The experimental design. (c) The corresponding real-world scenes of the example point visual information processing. (d) The visual information processing within the yellow box in Figure 4c, including the change in light intensity (from 0 to 2.56), the differentiation voltage, and the memristor state changes (1 representing high resistance state, 0 representing low resistance state). (e) Sudden movement of pedestrian at night. (f) Sudden movement of pedestrian during the day. (g) Sudden movement of pedestrian hiding behind a car, with background vehicles moving slowly. (h) The amplitude spectrum comparison between the compressed image and the differentiated image at the last moment in Figure 4e. (i) The amplitude spectrum comparison between the compressed image and the differentiated image at the last moment in Figure 4f. (j) The amplitude spectrum comparison between the compressed image and the differentiated image at the last moment in Figure 4g.

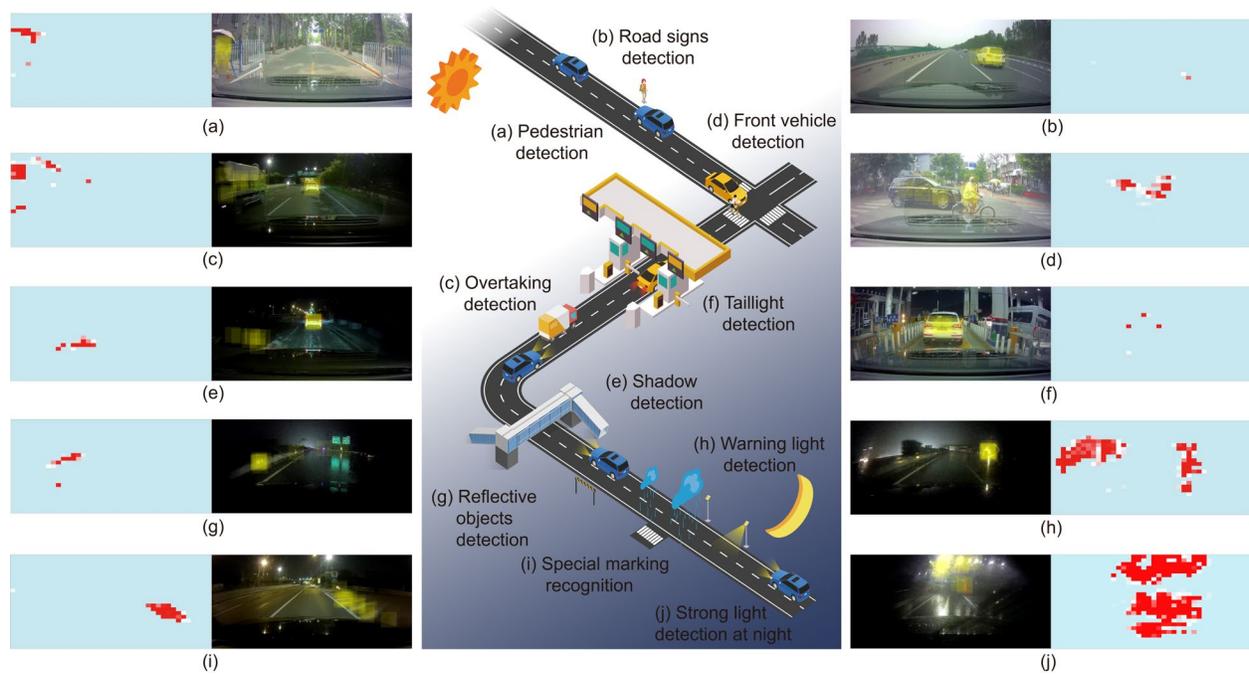

**Figure 5. Generalization capability of visual differentiation processing method in unstructured environments.** (a) Pedestrian detection. (b) Road signs detection. (c) Overtaking detection. (d) Front vehicle detection. (e) Shadow detection. (f) Taillight detection. (g) Reflective objects detection. (h) Warning light detection. (i) Special marking recognition. (j) Strong light detection at night.